\documentclass[journal=jacsat,manuscript=communication]{achemso}

\usepackage[version=3]{mhchem} 

\newcommand{\onlinecite}[1]{[\scalebox{1.3}[1.3]{\raisebox{-0.80ex}{\cite{#1}}}]}

\author{Shiyou Chen}
\affiliation[JCAP, LBNL]{Joint Center for Artificial Photosynthesis
(JCAP), Lawrence Berkeley National Laboratory, Berkeley, CA 94720,
United States} \alsoaffiliation[ECNU]{Key Laboratory of Polar
Materials and Devices (MOE), East China Normal University, Shanghai
200241, China} \email{shiyouchen@lbl.gov}

\author{Lin-Wang Wang}
\affiliation[JCAP, LBNL]{Joint Center for Artificial Photosynthesis
(JCAP), Lawrence Berkeley National Laboratory, Berkeley, CA 94720,
United States} \email{lwwang@lbl.gov}

\title[\texttt{achemso} demonstration]
{Thermodynamic Oxidation and Reduction Potentials of Photocatalytic
Semiconductors in Aqueous Solution}

\begin{document}
\begin{abstract}
We introduce an approach to calculate the thermodynamic oxidation
and reduction potentials of semiconductors in aqueous solution. By
combining a newly-developed ab initio calculation for compound
formation energy and band alignment with electrochemistry
experimental data, this approach can be used to predict the
stability of almost any compound semiconductor in aqueous solution.
30 photocatalytic semiconductors have been studied, and a graph (a
simplified Pourbaix diagram) showing their valence/conduction band
levels and oxidation/reduction potentials is produced. Based on this
graph, we have studied the stabilities and trends against the
oxidative and reductive photocorrosion for compound semiconductors.
We found that, only metal oxides can be thermodynamically stable
when used as the n-type photoanodes. All the non-oxides are unstable
due to easy oxidation by the photo-generated holes, but they can be
resistant to the reduction by electrons, thus stable as the p-type
photocathodes.

\end{abstract}

\begin{figure}
\scalebox{1.0}{\includegraphics{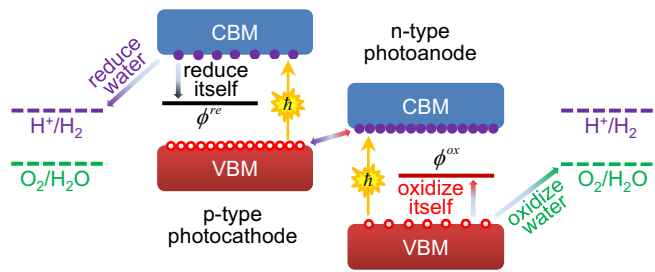}} \caption{A schematic plot
of the band alignment of the p-type photocathode and n-type
photoanode semiconductors relative to the water redox potentials in
the Z-scheme. $\phi^{ox}$ shows the oxidation potential of the
photoanode in aqueous solution, and $\phi^{re}$ shows the reduction
potential of the photocathode.} \label{sch:fig1}
\end{figure}

One key issue in the research of photocatalytic water splitting is to search for semiconductor photoelectrodes
 that can absorb the visible light and drive the hydrogen (oxygen) evolution reaction using the photo-generated electrons (holes).\cite{walter-6446-2010,cook-6474-2010,chen-6503-2010,hwang-410-2009}
This requires the semiconductors to have proper band alignment relative to the water redox potentials, \textit{e.g.} the conduction band minimum (CBM) of the p-type photocathode should
be higher (more negative in potential) than the water reduction
potential H$^+$/H$_2$, and the valence band maximum (VBM) of the n-type photoanode lower (more positive in potential) than the water oxidation potential O$_2$/H$_2$O,
as shown in the band alignment plot (\ref{sch:fig1}) for the Z-scheme water splitting system.\cite{gratzel-338-2001,mills-1-1997,gai-036402-2009,reece-6056-2011,maeda-3057-2011}
Given the importance of the band alignment, it is usually taken as
a screening condition in the search and design of new photocatalytic semiconductors. However, this condition is not sufficient, and one important issue,
which has attracted increasing attention but has not been well studied,
is how to evaluate and enhance the stability of semiconductors in the
aqueous solution.\cite{chen-6503-2010,chen-539-2011,bak-991-2002}

Resistance to the photo-induced corrosion (degradation
or decomposition) under illumination is a critical condition for the photocathode and photoanode materials.
It is due to this condition which makes the photocatalytic water splitting a much more challenging problem
than the photovoltaic.
A compound
semiconductor MX (\textit{e.g.} M=Zn, Ga, Ti, X=S, N, O$_2$ in ZnS, GaN,
TiO$_2$ respectively) used as the n-type photoanode may have a VBM
lower than the O$_2$/H$_2$O
oxidation potential, but the photo-generated holes ($h^+$) may
oxidize the semiconductor first, rather than the water, making the
compound MX decomposed through this reaction,\cite{gerischer-1422-1978,bak-991-2002}
\begin{equation}\label{e-1}
MX+zh^++solv \rightleftharpoons M^{z+}\bullet solv +X ~~(oxidization)
\end{equation}
Similarly the photo-generated electrons ($e^-$) in the p-type
photocathode MX may reduce itself rather than the water through the
following reaction,
\begin{equation}\label{e-2}
MX+ze^-+solv \rightleftharpoons M +X^{z-}\bullet solv ~~(reduction)
\end{equation}
The above reactions define two potentials for the hole and electron respectively. When the electron (hole) Fermi energy
equals that potential, the reactions are in equilibrium, \textit{i.e.} the Gibbs free energy
change equals zero, and when the photo generated electron (hole)
quasi Fermi energy is higher (lower) than that potential, the
reaction will occur, hence the semiconductor will be corroded. These potentials are called the thermodynamic reduction
potential $\phi^{re}$ and oxidization potential
$\phi^{ox}$ of the semiconductor.

\begin{figure}
\scalebox{0.5}{  \includegraphics{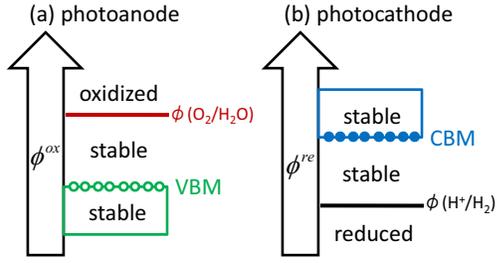}} \caption{The stability
change of the photoanode (a) as its oxidation potential $\phi^{ox}$
shifts up from below the VBM to above $\phi$(O$_2$/H$_2$O), and of
the photocathode (b) as its reduction potential $\phi^{re}$ shifts
down from above the CBM to below $\phi$(H$^+$/H$_2$).}
  \label{sch:fig2}
\end{figure}

Since the photoanode (photocathode) is in contact with
the aqueous solution, the photo-generated holes in the valence band
(electrons in the conduction band) can oxide (reduce) either the
water or itself, as shown in \ref{sch:fig2}. Whether the semiconductor is resistant
to the photocorrosion depends on the alignment of $\phi^{ox}$
relative to $\phi$(O$_2$/H$_2$O) for the photoanode, and $\phi^{re}$
relative to $\phi$(H$^+$/H$_2$) for the photocathode, as shown in
\ref{sch:fig2}. Generally speaking, a semiconductor is stable with respect to the hole oxidation if its $\phi^{ox}$ is lower than either $\phi$(O$_2$/H$_2$O) or its VBM,
and is stable with respect to the electron reduction if its $\phi^{re}$ is higher than either $\phi$(H$^+$/H$_2$) or its CBM.

Experimentally, $\phi^{ox}$ and $\phi^{re}$ can be derived from the Pourbaix diagram,\cite{pourbaix-1974,park-67-1979} but the diagram is not available for many novel photoelectrode semiconductors,
and it is not easy to measure it.
Thirty years ago, $\phi^{ox}$ and $\phi^{re}$ of several binary
semiconductors had been calculated by
Gerischer\cite{gerischer-133-1977,gerischer-1422-1978} (where the
labels $_p\epsilon_{decomp}$ and $_n\epsilon_{decomp}$ correspond to
$\phi^{ox}$ and $\phi^{re}$ respectively) and Bard and
Wrighton\cite{bard-1706-1977}. Park and Barber have also calculated the full Pourbaix diagram for a few simple semiconductor
compounds \cite{park-67-1979}.
These calculated $\phi^{ox}$ and
$\phi^{re}$  are widely cited in
literatures to explain the corrosion of semiconductors in aqueous
solution.\cite{morrison-1980,khase-3335-1998,bak-991-2002} As
proposed by Lewis and coauthors in their review
paper,\cite{walter-6446-2010} the $\phi^{ox}$ and $\phi^{re}$ should
be taken into account when choosing the candidate
photoelectrodes, in additional to the optimal band gap and band edge
alignment. Because the calculation of $\phi^{ox}$ and $\phi^{re}$ requires the Gibbs free energy of all the reactants and products in the related reactions, which are
not available for some semiconductors (especially for multinary ones),
$\phi^{ox}$ and $\phi^{re}$ of only a few simple semiconductors had been reported, and no general trend had been studied systematically. Recently new ab initio methods based
on the density functional theory have been developed to calculate the formation energies of semiconductors with high accuracy\cite{stevanovic-2011}, and also more experimental electrochemistry
data can been found in the latest handbooks,\cite{speight-2005,lide-2003} thus a more extended study of $\phi^{ox}$ and $\phi^{re}$ for almost all semiconductors becomes possible.
To facilitate the future choice of candidate
photocatalytic materials, we here report the
$\phi^{ox}$ and $\phi^{re}$ for more than 30
semiconductors as well as their band alignment relative to the normal hydrogen electrode (NHE) potential.  Based on these results, we will discuss the trend in the
oxidative/reductive stabilities of a series of metal oxides, II-VI and III-V
 related semiconductors.

To start with, we will take ZnS as an example and introduce how we
calculate its $\phi^{ox}$ and $\phi^{re}$. The specific
exemplification of Eqs.(1-2) for ZnS are,
\begin{equation}\label{e-3}
ZnS+2h^++H_2O \rightleftharpoons ZnO+S+2H^+ ~~(oxidization)
\end{equation}
\begin{equation}\label{e-4}
ZnS+2e^-+2H^+ \rightleftharpoons Zn+H_2S  ~~ (reduction)
\end{equation}
which define the $\phi^{ox}$ and $\phi^{re}$ of ZnS
respectively, \textit{i.e.} when the chemical potentials (Fermi energies) of the holes and electrons are equal to $\phi^{ox}$ and $\phi^{re}$ respectively,
the Gibbs free energy changes of Eq.(3) and (4) are zero, thus,
\begin{equation}\label{e-5}
\phi^{ox} = [G(ZnO)+G(S)+2G(H^+)-G(ZnS)-G(H_2O)]/2eF
\end{equation}
\begin{equation}\label{e-6}
\phi^{re}=-[G(Zn)+G(H_2S)-G(ZnS)-2G(H^+)]/2eF
\end{equation}
where G(X) stands for the Gibbs free energy of X at the standard state, \textit{e} is the elementary charge and \textit{F} is the Faraday constant.
To reference the potentials to the hydrogen reduction potential $\phi$(H$^+$/H$_2$), the following half reactions can be used:
\begin{equation}\label{e-7}
H_2+2h^+ \rightleftharpoons 2H^+
\end{equation}
\begin{equation}\label{e-8}
2H^++2e^- \rightleftharpoons H_2
\end{equation}
Here the chemical potentials of both the electron and hole are equal to $\phi$(H$^+$/H$_2$), thus $h^+$ is equivalent to $-e^-$, and
the above two equations are equivalent. $\phi$(H$^+$/H$_2$) can be written as
\begin{equation}\label{e-9}
\phi(H^+/H_2)=[2G(H^+)-G(H_2)]/2eF
\end{equation}
Now, we can add the reverse of Eq.(7) [exchange the left and right hand side] to Eq.(3), then get the sum reaction,
\begin{equation}\label{e-10}
ZnS+H_2O \rightarrow ZnO+S+H_2 ~~ (oxidization)
\end{equation}
and add the reverse of Eq.(8) to Eq.(4), then get
\begin{equation}\label{e-11}
ZnS+H_2 \rightarrow Zn+H_2S  ~~ (reduction)
\end{equation}
Note, to make $h^+$ in Eq. (3) and (7) and $e^-$ in (4) and (8) equivalent, it is assumed that their chemical potentials are the same as $\phi$(H$^+$/H$_2$),
so now the Gibbs free energy changes ($\Delta G$) of Eq. (3) and (4) are not zero, and they are equal to $\Delta G$ of the reactions (10) and (11) respectively,
thus the sign $\rightarrow$ is used rather than $\rightleftharpoons$. $\Delta G$ of the reactions (10) and (11) can be calculated directly,
\begin{equation}\label{e-12}
\Delta G(10)=G(ZnO)+G(S)+G(H_2)-G(ZnS)-G(H_2O)
\end{equation}
\begin{equation}\label{e-13}
\Delta G(11)=G(Zn)+G(H_2S)-G(ZnS)-G(H_2)
\end{equation}
From Eq. (5), (6), (9), (11) and (12), we have
\begin{equation}\label{e-14}
\phi^{ox}=\Delta G(10)/2eF+\phi(H^+/H_2)
\end{equation}
\begin{equation}\label{e-15}
\phi^{re}=-\Delta G(11)/2eF+\phi(H^+/H_2)
\end{equation}
Since the NHE potential is $\phi$(H$^+$/H$_2$) at pH=0, $\phi^{ox}$ and $\phi^{re}$ relative to NHE can be calculated using Eq. (14) and (15).
The key of the procedure is to find the plausible oxidation and reduction reactions, such as (10) and (11), and get their $\Delta G$, then the
relevant $\phi^{ox}$ and $\phi^{re}$ can be calculated. Several plausible reactions can be tried, then the lowest $\phi^{re}$ and highest $\phi^{ox}$ should
be used as the true reduction and oxidization potentials. Thus, in a sense, what we get is a simplified Pourbaix diagram, only the
potentials limiting the photo-corrosions are reported here.
In the supplement materials, we listed the reactions we
considered in determining $\phi^{ox}$ and $\phi^{re}$ for different semiconductors. It should be mentioned that there is no guarantee that we have
enumerated all the possible reactions, thus strictly speaking, the
potentials reported here should be considered as a higher limit for $\phi^{re}$ and a lower limit for $\phi^{ox}$.
Nevertheless, in practice, we believe the results we obtained are very close to the true oxidation/reduction potentials.

\begin{figure*}
\scalebox{0.6}{  \includegraphics{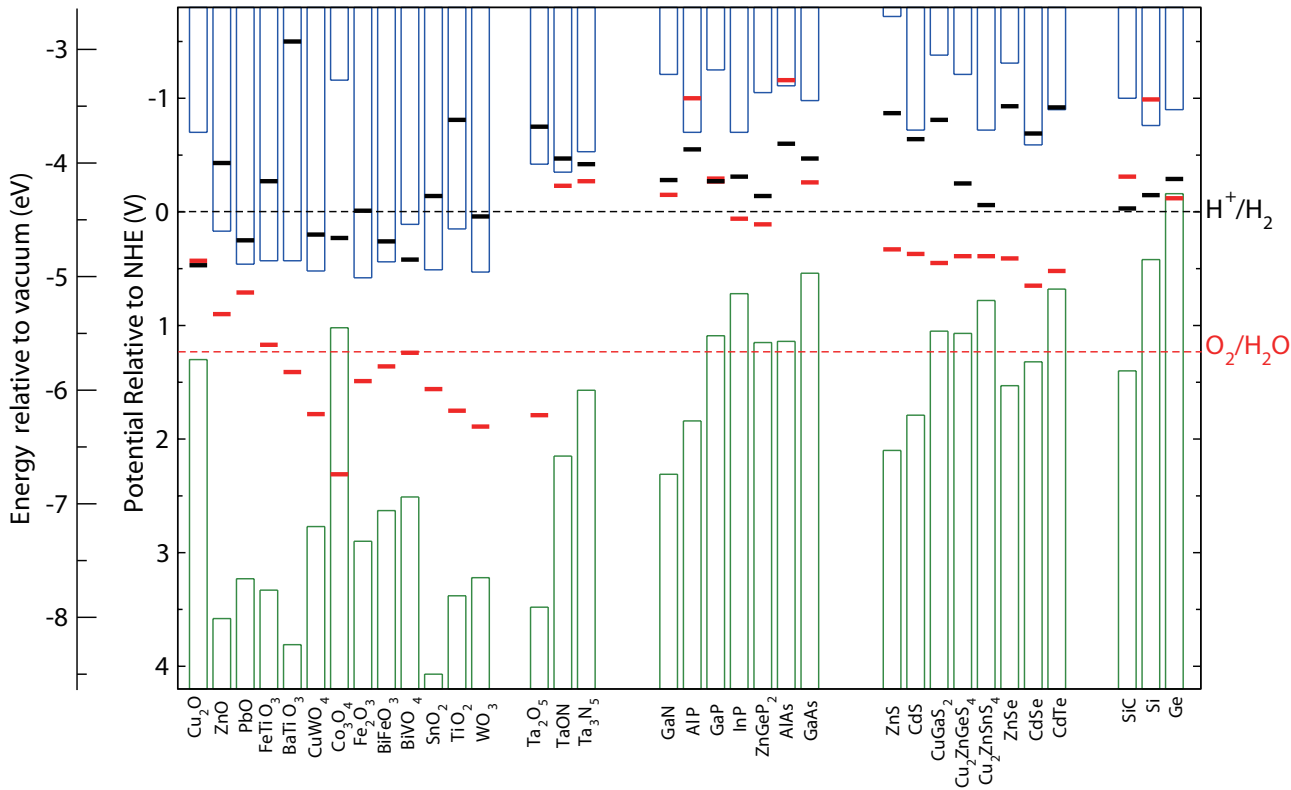}} \caption{The calculated
oxidation potential $\phi^{ox}$ (red bars) and reduction potential
$\phi^{re}$ (black bars) relative to the NHE and vacuum level for a
series of semiconductors in solution at pH=0. The water redox
potentials $\phi$(O$_2$/H$_2$O) and $\phi$(H$^+$/H$_2$) (dashed
lines), and the valence (green columns) and conduction (blue
columns) band edge positions at pH=0 are also
 plotted. The alignment for these potentials at different pH values (a simplified Pourbaix diagram) can be derived according to the relations described in the text, such as the Nernstian relation.}
  \label{sch:fig3}
\end{figure*}

\ref{sch:fig3} shows our calculated $\phi^{ox}$ and $\phi^{re}$
relative to NHE, the water redox potentials $\phi$(O$_2$/H$_2$O) and $\phi$(H$^+$/H$_2$), and the valence and conduction band edges for five selected classes of semiconductors
at pH=0. It is well-known that
$\phi$(O$_2$/H$_2$O) and $\phi$(H$^+$/H$_2$) depend on the pH value according to Nernstian relation,\cite{xu-543-2000} \textit{i.e.} shifting up by
-0.059 V as the pH increases by 1. The dependence of $\phi^{ox}$ and $\phi^{re}$ on the pH value is determined by the specific reactions:
when the sum reactions such as Eq. (10) and (11) are irrelevant to H$^+$ or OH$^-$, their $\Delta G$ is fixed, then $\phi^{ox}$ and $\phi^{re}$ shift together with $\phi$(H$^+$/H$_2$) as described
by Eq. (14) and (15) and also follow the Nernstian relation; when the reactions are relevant to H$^+$ or OH$^-$, the dependence is different, at variance to the Nernstian relation.
Most of $\phi^{ox}$ and $\phi^{re}$ plotted in \ref{sch:fig3} follow the Nernstian relation, similar as $\phi$(H$^+$/H$_2$), and
the exceptions include: (i) all oxides have fixed $\phi^{ox}$ with respect to pH change. (ii) $\phi^{ox}$ of CuGaS$_2$, Cu$_2$ZnGeS$_4$ and
Cu$_2$ZnSnS$_4$ shift up by -0.044 V (instead of -0.059 V) as pH increases by 1.  With these relations clear, the $\phi^{ox}$ and $\phi^{re}$
relative to NHE or $\phi$(O$_2$/H$_2$O) and $\phi$(H$^+$/H$_2$) at different pH can be plotted (as a simplified Pourbaix diagram).

The valence and conduction band edges in \ref{sch:fig3}  are collected from Ref.
\onlinecite{bak-991-2002} for most metal oxides and from Ref.
\onlinecite{walle-626-2003,chun-1798-2003} for non-oxides. For the metal oxides, the band
edge position relative to NHE is reported to depend on the pH value
of the solution through the same Nernstian relation as for the water
redox potentials.\cite{xu-543-2000} For the
non-oxides, the relation between the band edge position and the pH
value is so far not clear.\cite{xu-543-2000}
Ref. \onlinecite{walle-626-2003} derived the band edge positions of the
group IV, III-V and II-VI semiconductors relative to NHE based on
the calculated band offsets among semiconductors, assuming that
the conduction band edge of Si corresponds to its electron affinity
(about -3.7 eV relative to vacuum level) and NHE is -4.44 eV
relative to vacuum level. This implicitly assumes that the band edge
positions for these materials do not depend on the pH value. The band offsets of
almost all group IV, III-V and II-VI related semiconductors had
been calculated using the ab initio methods with reasonable consistence compared to
experiments,\cite{li-212109-2009} and the similar calculation
procedures can be used to estimate the band edge positions of other
semiconductor whose experimental values are
unavailable.\cite{chen-125201-2010}.

Compared to the calculation performed by Gerischer \emph{et al.}\cite{gerischer-133-1977,gerischer-1422-1978},
Bard and Wrighton \cite{bard-1706-1977}, and
Park and Barber \cite{park-67-1979} 30
years ago, our current work benefits from the progresses in both experiment and ab
initio calculations. It is the combination of these two
progresses which allow us to calculate $\phi^{ox}$ and $\phi^{re}$
for almost any given semiconductors, no matter binary or multinary
compounds, as shown
in \ref{sch:fig3}. (i) Experimentally,  the Gibbs free energy data in handbooks like Ref.
\onlinecite{speight-2005,lide-2003} covers much more chemical compounds with higher accuracy compared to those published 30
years ago. (ii) If the experimental Gibbs free energy (formation
enthalpy plus entropy contribution) for a compound in the proposed
reaction is not reported in those handbooks, mostly for some novel compound semiconductors, its
formation energy can be calculated using modern ab
initio methods. The Gibbs free energy of that compound  is then calculated from the formation energy plus the Gibbs free energies
of the elementary compounds which are all known experimentally.  For example, in this work, we have calculated the
formation energies of Ta$_2$N$_5$, TaON, ZnGeP$_2$, Cu$_2$ZnGeS$_4$,
Cu$_2$ZnSnS$_4$, etc. using a newly developed approach
\cite{stevanovic-2011} based on the density functional theory. This
approach reduces the formation energy error from 0.25 eV/atom to
less than 0.05 eV/atom for the compound
semiconductors\cite{stevanovic-2011}. As a result, if a compound in
the reactions has 4 atoms per formula unit, the errors of the
calculated $\phi^{ox}$ and $\phi^{re}$  are reduced from 1 V to 0.2
V (assuming one electron is transfered in the reaction). Note that the entropy
contribution in the formation energy is neglected for those
semiconductors with no experimental data, but for crystals this
contribution is small and we estimate it causes
errors less than 0.2 V in $\phi^{ox}$ and $\phi^{re}$. In the
supplement materials, we listed how we get the Gibbs free energy of
all the solid compounds which are not available in Ref.
\onlinecite{speight-2005,lide-2003}.

In the following we will discuss the trends in \ref{sch:fig3} and
their influence on the design of photocatalytic system.

(i) For the oxidation potential, most metal oxides have $\phi^{ox}$
lower than $\phi$(O$_2$/H$_2$O) at pH=0,
indicating they are resistant to the hole oxidation and stable in
the solution, but four exceptions are also found, Cu$_2$O, ZnO,
PbO and FeTiO$_3$. The reasons for the easier oxidation of
Cu$_2$O, PbO and FeTiO$_3$ are obvious, since the cations in these
compounds are not at their highest valences and can be further oxidized to
higher valences (Cu$^+$ to Cu$^{2+}$, Pb$^{2+}$ to Pb$^{4+}$ and
Fe$^{2+}$ to Fe$^{3+}$).\cite{scanlon-246402-2011} Comparing Cu$_2$O with Cu$^+$ and CuWO$_4$
with Cu$^{2+}$, we can find that CuWO$_4$ has a much lower
$\phi^{ox}$ than Cu$_2$O, since both cations in CuWO$_4$ are in the
highest valence state. According to this simple trend, all the metal
oxide semiconductors at a not-highest valence state of cations are likely
susceptible to oxidation in water, and tend to be unstable (Note Co$_3$O$_4$ is an exception, Co has higher valence in Co$_2$O$_3$, but Co$_2$O$_3$ is not as stable as Co$_3$O$_4$, thus its $\phi^{ox}$ is quite low).
The easy oxidation of ZnO
is also unique considering that Zn is at its highest valence state
and its band gap is quite large with a very low valence band. Thus
the stability of ZnO cannot be predicted according to its valence
state of elements, or its valence band position.  This is related more to
its formation energy relative to other phases or compounds. Since $\phi^{ox}$ of the listed oxides does not shift with pH value, when
pH=7, the $\phi^{ox}$ of ZnO and FeTiO$_3$ fall below
$\phi$(O$_2$/H$_2$O), indicating that these compounds are stable under
illumination in the neutral or alkaline solution, which is
consistent with the observation of ZnO in solution: no
photocorrosion at pH=10 while complete decomposition at acid
pH=4.5.\cite{domenech-1123-1986}

(ii) $\phi^{ox}$ of all non-oxide semiconductors are higher than
$\phi$(O$_2$/H$_2$O), at least 0.5 V more at pH=0, and the alignment of $\phi^{ox}$
relative to $\phi$(O$_2$/H$_2$O) is not changed by pH value.
This indicates these non-oxide semiconductors are thermodynamically
unstable in aqueous solution and will be oxidized by the holes under
illumination. The reason is also simple: the anions such as
N$^{3-}$, P$^{3-}$, As$^{3-}$, S$^{2-}$, Se$^{2-}$, Te$^{2-}$ can
all be oxidized to neutral or positive valence states, \textit{e.g.}
N$^{3-}$ to N$_2$ or NO$_3^-$, and S$^{2-}$ to S or SO$_4^{2-}$.
It is interesting to note that the oxynitride TaON, which is between
the oxide and nitride and naively expected to combine the better
stability of the oxides and higher valence band (hence smaller band
gap) of the
nitrides,\cite{maeda-7851-2007,maeda-3057-2011,chen-2011} has the
oxidation potential close to the nitride Ta$_2$N$_5$, therefore it
does not inherit the good oxidative stability of the oxides. This
further indicates that although doping or alloying the oxide
semiconductors with weaker electron-negative anions can decrease the
band gap, they become less stable with respect to oxidation in water.

(iii) For the reduction potential $\phi^{re}$, all the non-oxide semiconductors
have higher $\phi^{re}$ than $\phi$(H$^+$/H$_2$) at pH=0, and thus are resistant to
the electron reduction under illumination, corresponding to the situation in \ref{sch:fig2}(b) with $\phi^{re}$ above $\phi$(H$^+$/H$_2$).
Since $\phi^{re}$ and $\phi$(H$^+$/H$_2$) change at the same rate with pH value, this situation is not influenced by the pH change.
Our results are consistent with the reported Pourbaix diagrams for CdS, CdSe, CdTe, GaP and GaAs,\cite{park-67-1979}
which clearly show that their reduction and oxidation potentials are negative (higher in Fig.3) relative to $\phi$(H$^+$/H$_2$) and $\phi$(O$_2$/H$_2$O) respectively at 0<pH<14.

(iv) For $\phi^{re}$ of the metal oxides,
 a few metal oxides have lower $\phi^{re}$ than $\phi$(H$^+$/H$_2$) at pH=0,
such as Cu$_2$O, PbO, CuWO$_4$, BiFeO$_3$, BiVO$_4$, WO$_3$ and
Co$_3$O$_4$, caused by the ease of reactions such as,
\begin{equation}\label{e-5}
Cu_2O+H_2 \rightarrow 2Cu+H_2O
\end{equation}
\begin{equation}\label{e-6}
WO_3+H_2 \rightarrow WO_2+H_2O
\end{equation}
In these compounds, the bonding between the metal cations and oxygen
anions are weaker than the H-O bonding in H$_2$O, so the above
reactions are energetically favorable. According to the experimental Pourbaix diagrams,\cite{pourbaix-1974} the WO$_3$/WO$_2$ and Cu$_2$O/Cu reduction potentials
are 0 V and 0.45 V relative to $\phi$(H$^+$/H$_2$) respectively, in good agreement with our calculated $\phi^{re}$.
Although PbO, CuWO$_4$, BiFeO$_3$ and WO$_3$ have lower $\phi^{re}$ than $\phi$(H$^+$/H$_2$),
they are still stable because their $\phi^{re}$ is above their CBM, and thus the conduction band electrons can reduce neither the water
nor the semiconductors (Note $\phi^{re}$, CBM and $\phi$(H$^+$/H$_2$)
change at the same rate with pH, so the stability analysis of these metal oxides is not changed by pH). However, Cu$_2$O, BiVO$_4$ and
Co$_3$O$_4$, have $\phi^{re}$ lower than both $\phi$(H$^+$/H$_2$) and CBM, corresponding to the alignment in \ref{sch:fig2}(b) with
$\phi^{re}$ below $\phi$(H$^+$/H$_2$), thus they will be reduced in the solution under illumination. Actually the reduction of photocatalytic p-type Co$_3$O$_4$ to inactive CoO had been observed experimentally.\cite{gasp-19362-2011}
All the other oxides listed in \ref{sch:fig3} have $\phi^{re}$ higher than $\phi$(H$^+$/H$_2$),
and thus are stable against reduction.

According to the above trends, one can find a series of metal oxides
which are both resistant to the hole oxidation and the electron
reduction and thus stable in the solution. Due to the lower CBM than
$\phi$(H$^+$/H$_2$), most oxides except Cu$_2$O, BiVO$_4$ and Co$_3$O$_4$
can not be used as the photocathode, but can be used as the photoanode given their low VBM. Furthermore, the
low VBM also makes the p-type doping difficult and most metal oxides
are intrinsically n-type according to the doping-limit
rule,\cite{zhang-3192-1998} which states that a semiconductor is
difficult to be doped p-type if its valence band is too low and is
difficult to be doped n-type if its conduction band is too high.
Therefore the metal oxides with low valence band should be better
used as the n-type photoanodes.

For the p-type photocathode, one can easily find metal non-oxides
which have higher CBM than $\phi$(H$^+$/H$_2$) and are also resistant to
the electron reduction, thus can be
used the photocathode if doped to p-type. According to the doping
limit rule, the p-type doping in the non-oxides are
easier than in the oxides due to the higher VBM. One may worry about
the bad stability of the non-oxides with respect to the hole
oxidation. However, we should note that, for the p-type
photocathode, (i) the downward band bending at the
semiconductor/water interface prevent the majority holes from reacting
with water; (ii) the photo generated hole is expected to flow to the
connected anode quickly through majority carrier conduction (at
least when the device is working),\cite{kaly-987-1981} thus in this sense we can ignore whether it is
oxidization resistant. As a result, some of the III-V and II-V
semiconductors with suitable band gaps can be used as photocathodes,
despite the fact they might be prone to oxidization from the pure
thermodynamic point of view. The same can be said for n-type
photoanodes where we are mostly concerned about their resistance to
oxidization.

Finally, we want to mention that the results represented in Fig.3
considered only the thermodynamic resistance to the reductive and
oxidative decomposition, but the real decomposition also depends on
the specific kinetic processes. If the material is stable
thermodynamically against the decomposition process, this
material is stable regardless of the kinetic process. However, if
the material is unstable thermodynamically, there might still be
ways to make it stable kinetically. For example, an oxide layer, which forms as a result of oxidation, but stops itself after a certain thickness,
can be a good protective layer and make the material stable kinetically. For some p-type photocathodes,
such protective oxide layer can also serve as a hole blocker,
which prevents the recombination of the minority electron and the
majority hole, and thus can be essential to the efficiency of a photochemical cell.

Note, through out our study, we have not considered the possibility
of the semiconductor compound to be dissolved in the solvent in dark
condition. That can correspond to a reaction  where the valences of
cations and anions in the compound are not changed, but their total
Gibbs energy is smaller in the solvation form than as  in a solid
crystal structure. For example, this can happen to GaAs within some
pH range \cite{park-67-1979}. However, we do find some interesting
cases for some compounds where reactions in the dark will occur.
Comparing the oxidation potential $\phi^{ox}$ and reduction
potential $\phi^{re}$ in \ref{sch:fig3}, most compounds have
$\phi^{ox}$ lower than $\phi^{re}$, which is easy to understand
since the oxidation means the electrons are taken away from the low
valence band while the reduction means the electrons are added to
the high conduction band. However, several exceptions exist,
\textit{i.e.} Cu$_2$O, AlP, GaP, AlAs, Si and SiC have $\phi^{re}$
lower than $\phi^{ox}$. As a result, the electron-hole pair needed
for the reduction and oxidization reactions can be generated
spontaneously (they cost negative energy), without the help of any
photon. This will cause these materials to decompose even in the
dark. For example, the following two reactions are used to calculate
the $\phi^{re}$ and $\phi^{ox}$ of AlP respectively,
\begin{equation}\label{e-7}
2AlP+3H_2 \rightarrow 2Al+2PH_3   ~~ (reduction)
\end{equation}
\begin{equation}\label{e-8}
2AlP+3H_2O \rightarrow Al_2O_3+2P+3H_2 ~~ (oxidization)
\end{equation}
giving that $\phi^{re}$ is lower than $\phi^{ox}$, which makes the two
reactions spontaneous, corresponding to
a sum reaction,
\begin{equation}\label{e-9}
4AlP+3H_2O \rightarrow 2Al+Al_2O_3+2P+2PH_3
\end{equation}
The calculated Gibbs free energy change of this reaction is
negative, indicating the reaction is exothermal, and thus AlP will
be decomposed in water even without any illumination.
But once again, a kinetic barrier (e.g., an oxide layer) can block the
otherwise thermodynamically plausible reaction.

In conclusion, we studied the thermodynamic oxidation and reduction
potentials for a series of photocatalytic semiconductors, and
plotted their alignment relative to the valence and conduction band
edges as well as the water redox potentials. According to these
potentials, we found only metal oxides can be thermodynamically
stable when used as the n-type photoanodes, and all the non-oxides
are unstable due to easy oxidation by photo generated holes.
This indicates although the oxides doped or alloyed by less electronegative
anions can have smaller band gaps, their stability is also degraded. On
the other hand, many non-oxides are resistant to the electron
reduction, and thus may be used as p-type photocathode provided the hole
oxidation is prevented in the working devices. The method
we used is universal and can be applied to evaluate the stability of
any semiconductors. We believe the presented stability diagram
(\ref{sch:fig3}) will be useful in guiding the search for stable
photocathode and photoanode materials.

\acknowledgement We thank Dr. Jian-Wei Sun, Le Chen, Joel W. Ager
and Fei-Fei Shi for helpful discussions, and Chris Barrett for
proof reading the manuscript. This material is based upon work
performed by the Joint Center for Artificial Photosynthesis, a DOE
Energy Innovation Hub, as follows: The calculations about the metal
oxides, and the group III-V and IV related semiconductors were
supported through the Office of Science of the U.S. Department of
Energy under Award No. DE-SC0004993, using the supercomputers in
NERSC and NCCS; The calculations about II-VI related semiconductors
were supported by the Computer Center of ECNU.



\providecommand{\refin}[1]{\\ \textbf{Referenced in:} #1}

\end{document}